\documentclass{PoS}

\usepackage{psfrag,epsfig,graphicx,graphics}

\def\pv{\vec{p}_t}
\def\dv{\vec{\Delta}_t}

\def\ar{\alpha_\rho}

\title{Exploring GPDs through the photoproduction of a $\gamma \rho$ pair}

\ShortTitle{Photoproduction of a $\gamma \rho$ pair}

\author{\speaker{R.~Boussarie}\\
        Institute of Nuclear Physics, Polish Academy of Sciences, Radzikowskiego 152, PL-31-342 Krakow, Poland\\
        E-mail: \email{renaud.boussarie@ifj.edu.pl}}

\author{B.~Pire\\
 Centre de Physique Th\'eorique, \'Ecole Polytechnique,
CNRS, Universit\'e Paris-Saclay, 91128 Palaiseau,     France\\
        E-mail: \email{bernard.pire@polytechnique.edu} }

\author{L.~Szymanowski\\
 National Centre for Nuclear Research (NCBJ), 00-681 Warsaw, Poland\\
E-mail: \email{lech.szymanowski@ncbj.gov.pl}}
 
 \author{S.~Wallon\\
 LPT, Universit{\'e} Paris-Sud, CNRS, Universit\'e Paris-Saclay, 91405, Orsay, France {\em \&} \\
UPMC Univ. Paris 06, facult\'e de physique, 4 place Jussieu, 75252 Paris Cedex 05, France 
\\
E-mail: \email{samuel.wallon@th.u-psud.fr}}

\abstract{We describe the  process $\gamma N \to \gamma\,\rho N'$  in the generalized Bjorken regime  where the  $\gamma\,\rho $ pair has a large invariant mass.  In the collinear QCD factorization framework, the amplitude gives access to both chiral-even and chiral-odd quark generalized parton distributions (GPDs), and is insensitive to gluon  GPDs. The separation of  longitudinally and transversely polarized $\rho$ meson production allows to distinguish chiral-even and chiral-odd contributions. Production rates are estimated in the kinematics of the near-future JLab~12-GeV experiments. 
}

\FullConference{XXV International Workshop on Deep-Inelastic Scattering and Related Subjects\\
		3-7 April 2017\\
		University of Birmingham, UK}

\begin{document}

\begin{figure}[h]

\psfrag{TH}{$\Large T_H$}
\psfrag{Pi}{$\pi$}
\psfrag{P1}{$\,\phi$}
\psfrag{P2}{$\,\phi$}
\psfrag{Phi}{$\,\phi$}
\psfrag{Rho}{$\rho$}
\psfrag{tp}{$t'$}
\psfrag{s}{$s$}
\psfrag{x1}{$\!\!\!\!\!\!x+\xi$}
\psfrag{x2}{$\!x-\xi$}
\psfrag{RhoT}{$\rho_T$}
\psfrag{t}{$t$}
\psfrag{N}{$N$}
\psfrag{Np}{$N'$}
\psfrag{M}{$M^2_{\gamma \rho}$}
\psfrag{GPD}{$\!GPD$}

\centerline{
\raisebox{1.6cm}{\includegraphics[width=14pc]{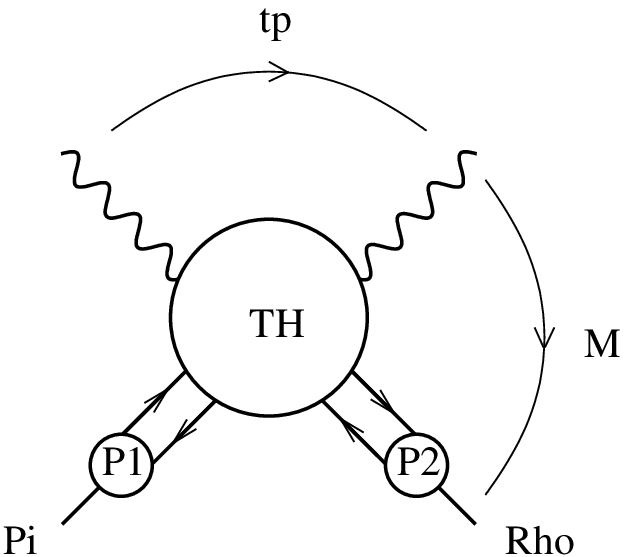}}~~~~~~~~~~~~~~
\psfrag{TH}{$\,\Large T_H$}
\includegraphics[width=14pc]{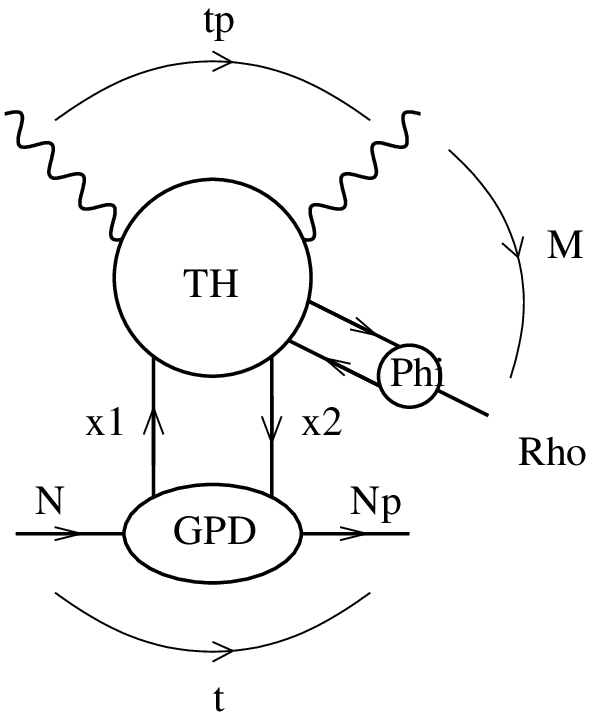}}

\caption{a) Factorization of the amplitude for the process $\gamma + \pi \rightarrow \gamma + \rho $ at large $s$ and fixed angle (i.e. fixed ratio $t'/s$); b) replacing the $\pi$ meson distribution amplitude by a nucleon generalized parton distribution  leads to the factorization of the amplitude  for $\gamma + N \rightarrow \gamma + \rho +N'$ at large $M_{\gamma\rho}^2$.}
\label{Fig:feyndiag}
\end{figure}

\section{Introduction}
We report here on our recent work on   exclusive photoproduction of a $\gamma\,\rho$ pair with a large invariant mass~\cite{BPSW}. In specific kinematics, this process may be described in the framework of collinear QCD factorization,  where the short distance part of the amplitude is calculated in a perturbative way. For this first study, this is done at first order in  the QCD coupling constant $\alpha_s$ with collinear kinematics and the long distance physics is encapsulated in leading twist hadronic matrix elements, namely the $\rho$ meson distribution amplitude (DA) and the nucleon generalized parton distributions (GPDs). 

  Exploring various exclusive processes in the generalized Bjorken regime is a mandatory step  to check the factorization hypothesis which allows to describe their amplitudes  in terms of GPDs with the final goal to explore the 3-dimensional structure of the nucleon, including its spin content~\cite{PRep}. Most of the theoretical and experimental effort has been up to now devoted to the analysis of  hard leptoproduction processes where a highly virtual photon probes the hadronic system, but the same experimental facilities produce intense real or quasi-real photon beams. Moreover, intense proton or nuclear high energy  beams like those of the LHC produce intense photon beams in the so-called ultra-peripheral  kinematics~\cite{UPC}. These beams open the way to the study of large invariant mass lepton pair~\cite{TCS} and hadron pair~\cite{pirho} exclusive production.

\section{Kinematics}
The process we study here
\begin{equation}
\gamma(q) + N(p_1) \rightarrow \gamma(k) + \rho^0(p_\rho,\varepsilon_\rho) + N'(p_2)\,,
\label{process1}
\end{equation}
may be described in the framework of collinear QCD by  first considering the factorization procedure of the wide angle Compton scattering on a meson~\cite{Lepage:1980fj} which amounts to write the leading twist
 amplitude for the process $\gamma + \pi \rightarrow \gamma + \rho $ shown in Fig.~\ref{Fig:feyndiag}a as the convolution of two mesonic DAs and a hard scattering subprocess amplitude $\gamma  +( q + \bar q) \rightarrow \gamma + (q + \bar q) $ with the meson states replaced by a collinear quark-antiquark pair. We then extract from the proof of factorization of 
exclusive meson electroproduction amplitude near the forward region~\cite{Collins:1996fb} the right to replace in Fig.~\ref{Fig:feyndiag}a the lower left meson DA by a $N \to N'$ GPD, and thus get Fig.~\ref{Fig:feyndiag}b. Such a factorization of a partonic amplitude requires to avoid the kinematical regions where a small momentum transfer is exchanged in the upper blob, namely small $t' =(k -q)^2$ or small $u'=(p_\rho-q)^2$, and the region where strong  final state interactions between the $\rho$ meson and the nucleon  are dominated by resonance effects, namely where the invariant mass $M^2_{\rho N'} = (p_\rho +p_{N'})^2$ is not large enough.

Introducing two light-cone vectors $p$ and $n$ (with $p \cdot n = \frac{s}2$), we write the particle momenta as
\begin{eqnarray}
\label{impini}
 p_1^\mu = (1+\xi)\,p^\mu + \frac{M^2}{s(1+\xi)}\,n^\mu~, \quad p_2^\mu = (1-\xi)\,p^\mu + \frac{M^2+\vec{\Delta}^2_t}{s(1-\xi)}n^\mu + \Delta^\mu_\bot\,, \quad q^\mu = n^\mu ~,
\end{eqnarray}
\begin{eqnarray}
\label{impfinc}
k^\mu = \alpha \, n^\mu + \frac{(\vec{p}_t-\vec\Delta_t/2)^2}{\alpha s}\,p^\mu + p_\bot^\mu -\frac{\Delta^\mu_\bot}{2}~, \quad \! \!
 p_\rho^\mu = \alpha_\rho \, n^\mu + \frac{(\vec{p}_t+\vec\Delta_t/2)^2+m^2_\rho}{\alpha_\rho s}\,p^\mu - p_\bot^\mu-\frac{\Delta^\mu_\bot}{2}\,,\quad \
\end{eqnarray}
with  
$M$, $m_\rho$ the masses of the nucleon and the $\rho$ meson. The  squared center-of-mass energy of the $\gamma$-N system is
then $
S_{\gamma N} = (q + p_1)^2 = (1+\xi)s + M^2$, while the small  squared transferred momentum is
$
t = (p_2 - p_1)^2 = -\frac{1+\xi}{1-\xi}\vec{\Delta}_t^2 -\frac{4\xi^2M^2}{1-\xi^2}$. The hard scale $M^2_{\gamma\rho}$ is the invariant squared mass of the $\gamma$ $\rho$ system.  In the generalized Bjorken limit, the approximate kinematics  allows to neglect  $\dv$ in front of $\pv$ as well as hadronic masses, leading to
\begin{eqnarray}
\label{skewness2}
M^2_{\gamma\rho} \approx  \frac{\vec{p}_t^2}{\alpha\bar{\alpha}} ~, ~
\ar \approx 1-\alpha \equiv \bar{\alpha} ~,~
\xi =  \frac{\tau}{2-\tau} ~,~\tau \approx 
\frac{M^2_{\gamma\rho}}{S_{\gamma N}-M^2}~,~
-t'  \approx  \bar\alpha\, M_{\gamma\rho}^2  ~,~ -u'  \approx  \alpha\, M_{\gamma\rho}^2 \,.\quad \,
\end{eqnarray}
It is interesting to note the analogy with the kinematics of timelike Compton scattering~\cite{TCS}. However, the more complex momentum flow of the present process leads to the coexistence of both timelike ($M_{\gamma\rho}^2$) and spacelike ($u' $) large scales, allowing a more complex analytic structure of the amplitude~\cite{MPSW}. 

\section{Ingredients}
One of the peculiar features of our process is its sensitivity  to both chiral-even and chiral-odd GPDs due to the chiral-even (resp. chiral-odd) character of the leading twist DA of $\rho_L$ (resp. $\rho_T$). Indeed, these twist 2 DAs are defined as
\begin{eqnarray}
\langle 0|\bar{u}(0)\gamma^\mu u(x)|\rho^0(p_\rho,\varepsilon_{\rho_L}) \rangle &=& \frac{1}{\sqrt{2}}p_\rho^\mu f_{\rho^0}\int_0^1dz\ e^{-izp_\rho\cdot x}\ \phi_{\parallel}(z), \\
\langle 0|\bar{u}(0)\sigma^{\mu\nu}u(x)|\rho^0(p_\rho,\varepsilon_{\rho_\pm}) \rangle &=& \frac{i}{\sqrt{2}}(\varepsilon^\mu_{\rho_\pm}\, p^\nu_\rho - \varepsilon^\nu_{\rho_\pm}\, p^\mu_\rho)f_\rho^\bot\int_0^1dz\ e^{-izp_\rho\cdot x}\ \phi_\bot(z),
\end{eqnarray}
where $\varepsilon^\mu_{\rho_\pm}$ is the $\rho$-meson transverse polarization and with $f_{\rho^0}$ = 216 MeV and $f_\rho^\bot$ = 160 MeV. 

As for the GPDs, they are  defined as usual~\cite{PRep}; in particular 
the transversity GPD of a quark $q$   is defined by:
\begin{eqnarray}
\langle p(p_2)|\, \bar{q}\left(-\frac{y}{2}\right)i\,\sigma^{+j} q \left(\frac{y}{2}\right)|p(p_1)\rangle = \int_{-1}^1dx\ e^{-\frac{i}{2}x(p_1^++p_2^+)y^-}\bar{u}(p_2)\, \left[i\,\sigma^{+j}H_T^{q}(x,\xi,t) +\dots
\right]u(p_1)\,,\nonumber \\ 
\end{eqnarray}
where $\dots$ denote the remaining three chiral-odd GPDs which contributions are omitted in the present analysis, in the small $\xi$ limit.
We parametrized the GPDs in terms of double distributions without including the quite arbitrary $D$ term. 

\section{The Scattering Amplitude}
The computation of the scattering amplitude of the process is straightforward at leading order in $\alpha_s$, although the number of Feynman diagrams is quite large. After a tensorial decomposition is applied, the integral with respect to the variable $z$ entering the meson DA is trivially performed in the case of a DA expanded in the basis of Gegenbauer polynomials. The integration with respect to the variable $x$ entering the GPDs  is then reduced to the numerical evaluation of a few building block integrals. Details can be found in the appendix of Ref.~\cite{BPSW}.
\psfrag{H}{\hspace{-1.5cm}\raisebox{-.6cm}{\scalebox{.9}{$M^2_{\gamma \rho}~({\rm GeV}^{2})$}}}
\psfrag{V}{\hspace{-.5cm}\raisebox{.3cm}{\scalebox{.9}{$\hspace{-.4cm}\displaystyle\frac{d\sigma_{even}}{d M^2_{\gamma\rho}}~({\rm nb} \cdot {\rm GeV}^{-2})$}}}
\begin{figure}[!h]
\begin{center}
\psfrag{T}{}
\vspace{.3cm}
\hspace{.2cm}\includegraphics[width=7.3cm]{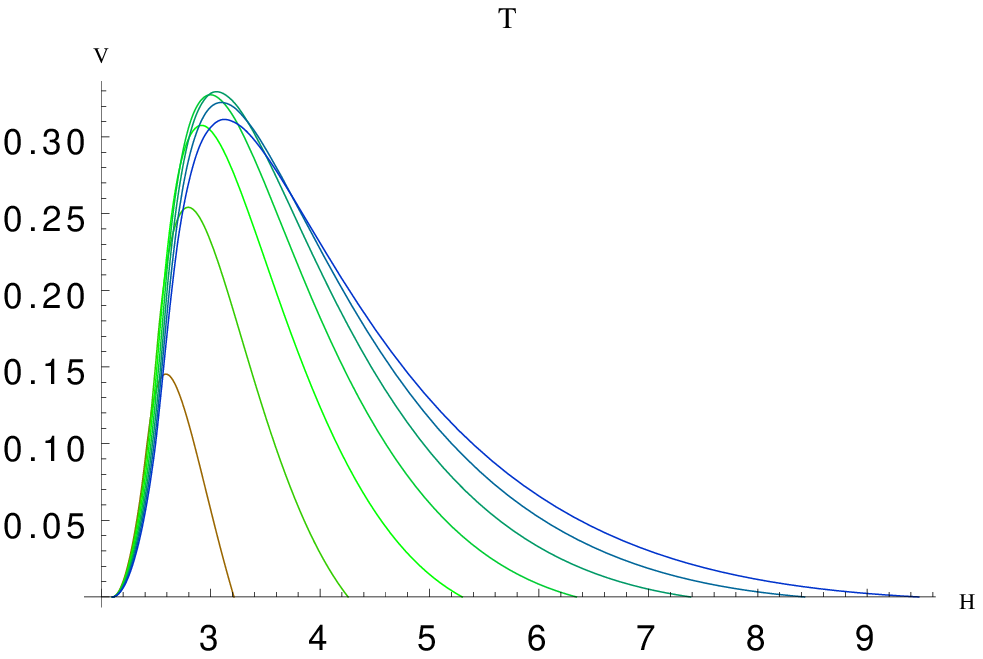}
\psfrag{T}{}
\hspace{0.1cm}\includegraphics[width=7.3cm]{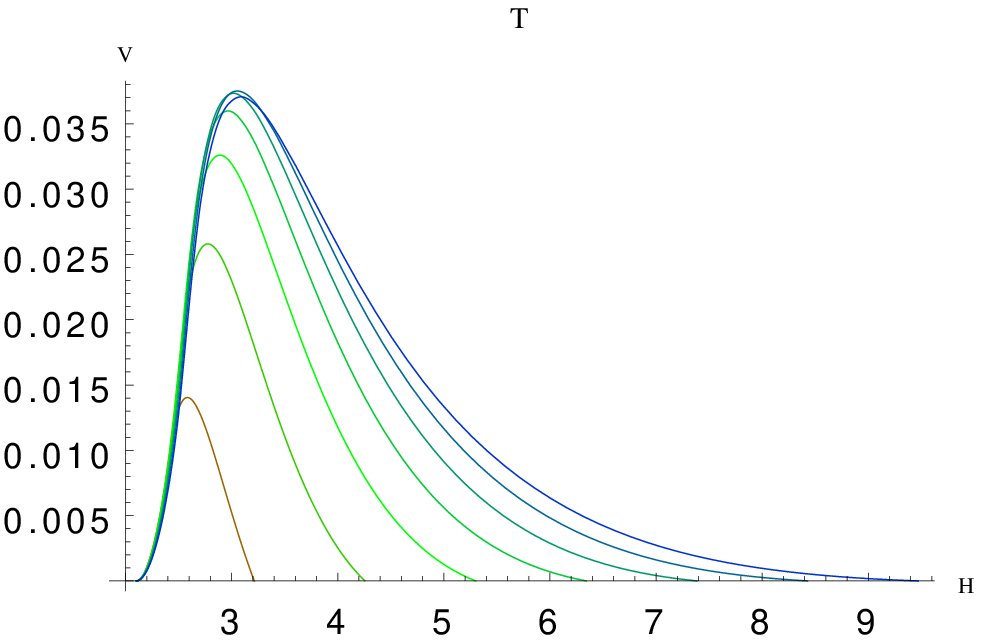}
\vspace{.2cm}
\caption{
DDDDifferential cross section $d\sigma/dM^2_{\gamma \rho}$ for a photon and a  longitudinally polarized $\rho$ meson production, on a proton (left) or neutron (right) target. The values of $S_{\gamma N}$ vary in the set 8, 10, 12, 14, 16, 18, 20 ${\rm GeV}^{2}.$ (from 8: left, brown to 20: right, blue), covering the JLab energy range.}
\label{Fig:dsigmaEVENdM2SgN8,10,12,14,16,18,20}
\end{center}
\end{figure}
\psfrag{H}{\hspace{-1.5cm}\raisebox{-.9cm}{\scalebox{.9}{$M^2_{\gamma \rho} ({\rm GeV}^{2})$}}}
\psfrag{V}{\hspace{-.5cm}\raisebox{.3cm}{\scalebox{.9}{$\hspace{-.7cm}\displaystyle\frac{d \sigma_{\rm odd}}{d M^2_{\gamma \rho}}~({\rm pb} \cdot {\rm GeV}^{-2})$}}}
\psfrag{T}{}
\begin{figure}[!h]
\begin{center}
\includegraphics[width=10cm]{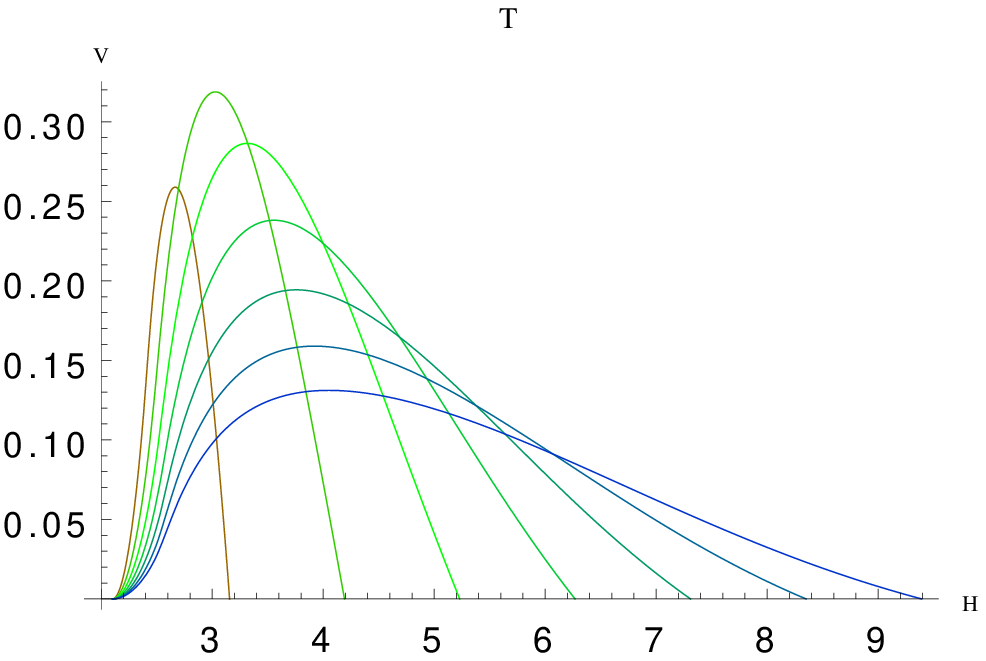}
\vspace{.4cm}

\caption{
Differential cross section $d\sigma/dM^2_{\gamma \rho}$ for a photon and a  transversally polarized $\rho$ meson production, on a proton  target. The values of $S_{\gamma N}$ vary in the set 8, 10, 12, 14, 16, 18, 20 ${\rm GeV}^{2}$ (from 8: left, brown to 20: right, blue), covering the JLab energy range.}
\label{Fig:dsigmaODDdM2SgN8,10,12,14,16,18,20}
\end{center}
\end{figure}

\section{Cross-sections}
The differential cross section as a function of $t$, $M^2_{\gamma\rho},$ $-u'$  reads
\begin{equation}
\label{difcrosec}
\left.\frac{d\sigma}{dt \,du' \, dM^2_{\gamma\rho}}\right|_{\ -t=(-t)_{min}} = \frac{|\mathcal{\overline{M}}|^2}{32S_{\gamma N}^2M^2_{\gamma\rho}(2\pi)^3}\,.
\end{equation}
By lack of space, we refer the interested reader to Ref.~\cite{BPSW} for a detailed analysis of this fully differential cross-section. To get an estimate of the total rate of events of interest for our analysis, we restrict here to  the $M^2_{\gamma\rho}$ dependence of the differential cross section integrated over $u'$ and $t$,
\begin{equation}
\label{difcrosec2}
\frac{d\sigma}{dM^2_{\gamma\rho}} = \int_{(-t)_{min}}^{(-t)_{max}} \ d(-t)\ \int_{(-u')_{min}}^{(-u')_{max}} \ d(-u') \ F^2_H(t)\times\left.\frac{d\sigma}{dt \, du' d M^2_{\gamma\rho}}\right|_{\ -t=(-t)_{min}} \,.
\end{equation}
The obtained differential cross sections for the longitudinal and transverse polarization cases, $d\sigma/dM^2_{\gamma \rho}$ are shown in Fig.~\ref{Fig:dsigmaEVENdM2SgN8,10,12,14,16,18,20} and in  Fig.~\ref{Fig:dsigmaODDdM2SgN8,10,12,14,16,18,20} for various values of $S_{\gamma N}$ covering the JLab-12 energy range. These cross sections show a maximum around $M^2_{\gamma \rho}\approx 3~$GeV$^2$, for most energy values. The order of magnitude of the cross sections are large enough for the measurement to seem feasible at JLab. Longitudinal $\rho$ production clearly dominates over the transverse $\rho$ production, at least with our models of the GPDs. To get a better access to the elusive transversity GPDs~\cite{transGPDno}, one may have to measure the off-diagonal spin matrix components $\rho_{10}$ which is linear in the transversity GPD and measurable through the angular dependence of the $\rho$ meson decay. 
\\

Let us note that to confirm the order of magnitude of our present study, the effect of next-to-leading-order corrections, using for example the method of Refs.~\cite{Nizic:1987sw_Duplancic:2006nv}, should be evaluated, as well as the effect of the renormalization/factorization scale fixing (taken here at fixed value) which should be done with care for exclusive processes~\cite{BLM-exclusive}. This is left for future studies.

\paragraph*{Acknowledgements.}
\noindent
 This work is partly supported by grant No 2015/17/B/ST2/01838 from the National Science Center in Poland,  by the French grant ANR PARTONS (Grant No. ANR-12-MONU-0008-01),  by the Labex P2IO and by the Polish-French collaboration agreements  Polonium and COPIN-IN2P3.

\end{document}